\documentclass[conference]{IEEEtran}
\IEEEoverridecommandlockouts

\usepackage{cite}
\usepackage{amsmath,amssymb,amsfonts}
\usepackage{algorithmic}
\usepackage{graphicx}
\usepackage{textcomp}
\usepackage{xcolor}
\usepackage{verbatim}
\usepackage{array}
\usepackage{url}
\usepackage{subfigure}
\usepackage{comment}
\usepackage{esdiff}
\usepackage[symbol]{footmisc}
\def\BibTeX{{\rm B\kern-.05em{\sc i\kern-.025em b}\kern-.08em
    T\kern-.1667em\lower.7ex\hbox{E}\kern-.125emX}}
    
\begin{document}

\title{Metastability-Resilient Synchronization FIFO for SFQ Logic\thanks{This work was supported by the Office of the Director of National Intelligence (ODNI), the Intelligence Advanced Research Projects Activity (IARPA), via the U.S. Army Research Office Grant W911NF-17-1-0120.}}

\author{\IEEEauthorblockN{Gourav Datta, Haolin Cong, Souvik Kundu, Peter A.~Beerel}
\IEEEauthorblockA{\textit{Ming Hsieh Department of Electrical and Computer Engineering} \\
\textit{University of Southern California}\\
Los Angeles, California 90089, USA \\
\{gdatta, haolinco, souvikku, pabeerel\}@usc.edu}
}

\maketitle

\begin{abstract}
Digital single-flux quantum (SFQ) technology promises to meet the  demands of ultra low power and high speed computing needed for future exascale supercomputing systems. The combination of ultra high clock frequencies, gate-level pipelines, and numerous sources of variability in SFQ circuits, however, make low-skew global clock distribution a challenge. This motivates the support of multiple independent clock domains and related clock domain crossing circuits that enable reliable communication across domains. Existing J-SIM simulation models indicate that setup violations can cause clock-to-Q increases of up to $100$\%. This paper first shows that naive SFQ clock domain crossing (CDC) first-in-first-out buffers (FIFOs) are vulnerable to these delay increases, motivating the need for more robust CDC FIFOs. Inspired by CMOS multi-flip-flop asynchronous FIFO synchronizers, we then propose a novel $1$-bit metastability-resilient SFQ CDC FIFO that simulations show delivers over a $1000$ reduction in logical error rate at $30$ GHz. Moreover, for a 10-stage FIFO, the Josephson junction (JJ) area of our proposed design is only $7.5$\% larger than the non-resilient counterpart. Finally, we propose design guidelines that define the minimal FIFO depth subject to both throughput and burstiness constraints.    
\end{abstract}

\section{Introduction}\label{intro}
While conventional CMOS-based technology is approaching its physical limits, SFQ \cite{likharev1991} circuits are gaining attention due to their potential for fast switching and low energy consumption. However, the combination of high variability \cite{bunyk50, parameters1995}, gate-level pipelining, and ultra-fast clock frequencies makes clock distribution and timing extremely challenging \cite{timingSpringer}. Some have argued that large-scale clock-trees will not be sufficiently robust \cite{likharev1991, timingSpringer}. Others have proposed asynchronous or hybrid clocking strategies\cite{maezawa1997,kameda1998,ito2005, gerber2007}, which are expensive or require significant manual customization.   

We note that, similar to CMOS, to ease the problem of global clock skew, SFQ circuits can be designed to operate with multiple independent smaller clock domains. 
To ensure safe transfer of data and control signals between these clock domains, efficient clock domain crossing (CDC) circuits are required. Naive CDC circuits, however, can exhibit timing violations, which can cause metastability in the receiver clock domain \cite{Metastability}. If metastability is not handled properly, it may cause problems ranging from transient errors to failure of the whole chip.
The remainder of the paper is organized as follows. Section \ref{sec:baseline_CDC} explains the vulnerability of the state-of-the-art CDC synchronizers in SFQ, explaining the notion of metastability. Section \ref{sec:FIFO_sync} then presents the proposed FIFO Synchronizer, presents the timing conditions for proper Synchronization, and proposes design guidelines for the minimum number of FIFO stages, subject to throughput and burstiness constraints. The simulation platform and results are then summarized in Section \ref{sec:sim}. And finally some conclusions are given in Section \ref{sec:conc}. 
\section{State-of-the-Art CDC Synchronizers}

\label{sec:baseline_CDC}
In CMOS, elastic pipelines like FIFOs are popular as circuits generating local clocks.
These circuits can be implemented in SFQ using a shift-register \cite{likharev1991}, where each block waits for the arrival of a SEND pulse signaling
that the receiver cell is reset and hence ready to accept new
data, as in Fig. \ref{FIFO_designs}(a). This pulse triggers a 
C-junction to produce the clock pulse for its adjacent block and requests it
to send its output signal to the receiver. The C-element acts as an unclocked AND gate and provides its output pulse as soon as
both its inputs have been fed by such pulses. The clock pulse is duplicated as the acknowledge signal necessary to set up the coincidence junction of the receiver and (after an appropriate
delay) as the SEND' signal for the next block. Thus, the whole data string will be eventually shifted towards the output by one. 
\begin{figure*}[!t]
\centering {
 \vspace{-.5cm}
\includegraphics[width=18.6cm]{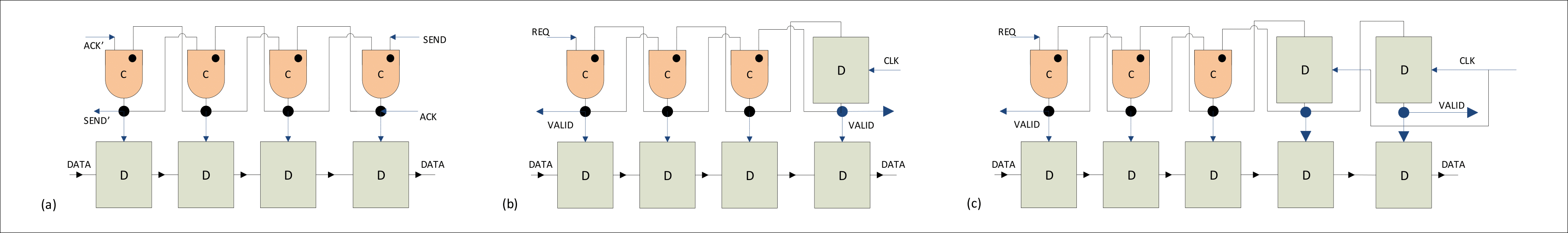}
\caption{(a) Shift Register based FIFO design, (b) Baseline FIFO design, (c) Proposed FIFO design.}
\label{FIFO_designs}
}
 \vspace{-.4cm}
\end{figure*}

A phase demodulator has been presented in \cite{ADC}, which can be used for synchronization between circuit blocks. However, the synchronizer can only support clocks of the same frequency.
Authors in \cite{FIFO} replace the C-element in the right-most stage of the structure proposed in \cite{likharev1991} by a destructive-read-out storage element, aka a D-element, to sample the output in the receiver clock domain, as shown in Fig. \ref{FIFO_designs}(a). They claim that this technique supports the synchronization of independent read and write clocks. However, if the final D-element experiences a large clock-to-Q delay, this claim can be invalidated due to a large increase in its' output delay. To demonstrate this concern, we simulated the clock-to-Q delay of a custom designed D-element in the SFQ5EE process as a function of the relative arrival time of the input and clock pulses. We define setup time as the time of arrival of the input pulse prior to the clock pulse which causes the clock-to-Q delay to rise 10\% over the nominal value. JSIM simulations indicate that the setup time for our D-element is 2.1ps. The nominal clock-to-Q delay is 8.1ps. It starts to increase when the input pulse is within 2ps of the clock pulse and increases drastically to 18ps when the data arrives 1.8ps before the clock. Finally at 1.7ps, the output is so late it arrives in the next clock cycle. The detailed results are shown in Fig. 1.
\vspace{-0.2cm}

\begin{figure}[!t]
        \centering {
        \hspace{-.6cm}
        \includegraphics[width = 0.8\linewidth]{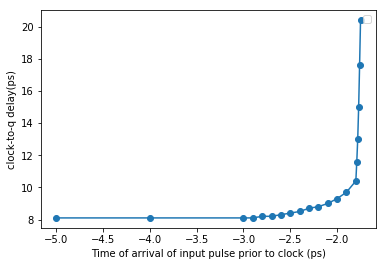}
        \vspace{-0.3cm}
        \caption{Variation of Clock-to-Q delay of a D Flip-flop with the relative arrival of the input pulse with respect to the clock pulse}
        \label{fig:DFF_setup}
        }
        \vspace{-0.6cm}
\end{figure}

\section{The Proposed FIFO Synchronizer: qCDC}
\label{sec:FIFO_sync}
\subsection{The Circuit}

We propose an efficient class of FIFO synchronizers for CDCs with reduced logical error rate. Our proposed design is inspired from the well-known two flip-flop asynchronous synchronizers used in CMOS \cite{ginosar}, where the output of the last flip-flop changes either one or two clock cycles after the input, but rarely becomes metastable. These synchronizers exchange the ‘analog’ uncertainty of metastability (continuous voltage levels changing over continuous time) for a simpler ‘digital’
uncertainty (discrete voltage levels switching only at
uncertain discrete time points) of whether the output switches one or two cycles later. However, being a pulse based logic, SFQ does not have analog voltage levels. The analog uncertainty in CMOS synchronizers is translated to the increased clock-to-Q delay in the right-most DRO of FIFO synchronizers (Fig. \ref{FIFO_designs}(b)).
Our proposed modification masks this clock-to-Q uncertainty and 
increases the likelihood of proper synchronization of 'VALID' and 'DATA' (Fig. \ref{FIFO_designs}(c)). However, such FIFOs still require 
appropriate stalling logic on the read side that is activated when the FIFO is empty and on the write side that is activated when the FIFO is full.

Our proposed design contains Muller C-elements, Josephson transmission lines (JTLs), splitters and DRO-elements, as illustrated in Fig. \ref{FIFO_designs}(c). The C-element requires its dotted conjugate where a tweak in the bias current distribution circuit results in the junction being initialized in a different state. This means that the first pulse on the dotted port results in an output pulse,
and it shall behave onward as an un-dotted C-junction. 
As illustrated in Fig. \ref{FIFO_designs}, the C-junction is implemented with three JJs, DRO gate with six, and the splitter requires three junctions. For all JJs, the value of Stewart-McCumber parameter has been kept two. Sometimes we require three-way fan out which is achieved by using a pair of splitters. Moreover, JTLs have been inserted between cells to provide electrical isolation as needed.

\vspace{-0.07cm}

\begin{figure}[!t]
\centering {
\includegraphics[width=8.6cm]{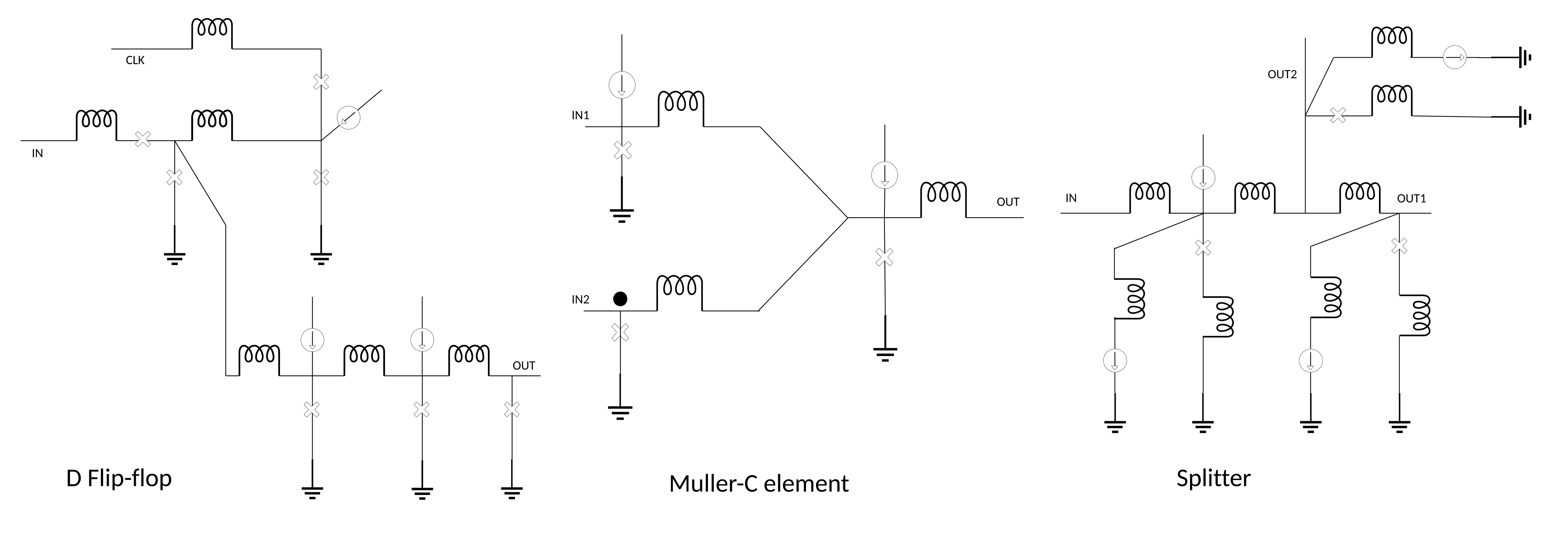}
 \vspace{-0.5cm}
\caption{Implementation of different cells of our proposed FIFO}
\label{FIFO_designs}
}
 \vspace{-.4cm}
\end{figure}
\subsection{Timing Analysis}
\begin{figure}[!t]
\centering
 \vspace{-.5cm}
\includegraphics[width = 0.65\linewidth]{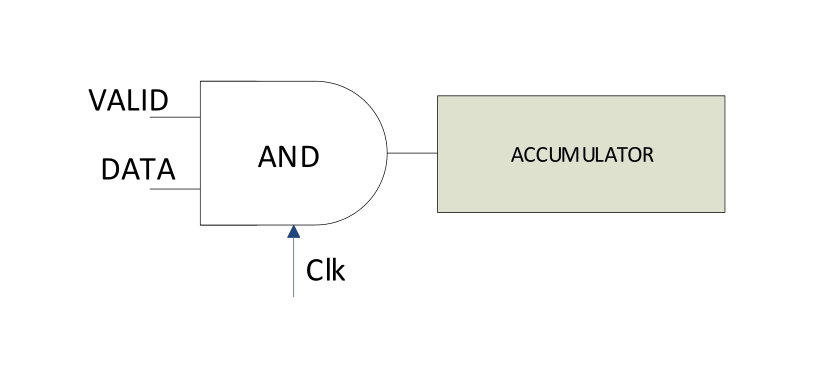}
\vspace{-0.7cm}
\caption{Circuit in the receiver domain to demonstrate timing failure for wrong synchronization of VALID and DATA signal}
\vspace{-0.5cm}
\label{simple_circuit}
\end{figure}

Our proposed FIFO helps to reduce the timing failure by enabling proper synchronization of data with the control signal (VALID). An increase in clock-to-Q delay of the top right D flip-flop in Fig. \ref{FIFO_designs}(b) delays the generation of VALID Signal. Since, VALID has to incur the delay of a JTL and a splitter, it can appear near the end of the receiver clock period. The output data faces yet an additional JTL and clock-to-Q delay, which might make it arrive in the next receiver clock cycle, especially if the clock frequency is relatively high. VALID and DATA, if appears in different clock cycles can cause a timing error in any subsequent data which is expecting VALID and DATA to be synchronized.
Consider for example, the simple accumulator circuit illustrated in Fig. \ref{simple_circuit}. Consider the erroneous case of VALID sometimes occurring a cycle before DATA. Hence, the output of the accumulator will under count the number of DATA signals.  

Interestingly, if the clock-to-Q delay of the right D-element is very high, the VALID signal may also appear in the next clock cycle. In this case the output data and VALID signals are correctly synchronized (since both of them appears in the same cycle) but one clock cycle later than expected. Hence, there is a critical timing window, before the arrival of the clock pulse, in which the appearance of VALID causes a timing failure.  Logical Error Rate (LER) is proportional to the ratio of this timing window to the entire clock period.  

Our design addresses this problem by capturing the output of the D-element twice. Even if the clock-to-Q delay of the first D flip-flop falls within the dangerous timing window, the next D-element can come to the rescue. 
Thus, the last D-element more likely has its nominal clock-to-Q delay which helps VALID and DATA appear in the same clock cycle. 

Extending our proposed design with two DRO synchronizers yields higher {\em mean time between failure} (MTBF), but at the cost of higher area and latency. Adding more than two DROs further improves the performance (MTBF) but exhibits diminishing returns in terms of LER reduction.
\vspace{-0.2cm}

\subsection{Determining the Number of Stages}
The FIFO design has a peak throughput $1/P$\cite{beerel_book} constrained by the delay of one cycle of reading/writing to the FIFO and this throughput is independent of the number of FIFO stages. When run at lower than this throughput, referred to as the average throughput $T$, however, the FIFO can contain a range of tokens referred to as the {\em dynamic slack} that is also a function of the number of FIFO stages $N$ \cite[Chap. 4]{beerel_book}: 
\begin{IEEEeqnarray}{c}\label{eq:slack}
D=N(1-TP)  \IEEEyesnumber
\end{IEEEeqnarray}
This means that the FIFO can tolerate short bursts of reads faster than $T$, buffering the extra tokens within the FIFO, and the larger the FIFO size $N$, the longer such bursts can be.

\section{Simulation Platform and Results}
\label{sec:sim}
We have performed JSIM simulations for different read clock frequencies and different instances of arrival of the input SFQ pulse relative to the clock pulse to verify proper synchronization of data in the read clock domain. Simulations show that our proposed 2-DRO synchronizer reduce the LER (explained in Section \ref{sec:FIFO_sync}) of data in the read domain by a factor of $1000$ over the baseline FIFO \cite{FIFO}. Moreover, for a 10-stage FIFO, the JJ-area of our proposed design is only 7.5\% higher than the naive counterpart. Detailed results, comparing the synchronization successes/failures of the baseline and proposed designs for different arrival times of the input pulse and different read clock frequencies have been shown in Table \ref{tab1}.\footnote{Note that '-' means the signal appears on the next clock cycle.} Comparison of JJ-areas of the two designs have been presented in Table \ref{tab2}. The latency of our proposed synchronizer exceeds the non-resilient one by only one clock cycle.

\begin{table}
\vspace{-0.5cm}
\caption{Comparison of Performance of the two FIFOs}
\vspace{-0.1cm}
\centering
\tabcolsep=0.11cm
\small
\resizebox{\columnwidth}{40pt}{
\begin{tabular}{|c|c|c|c|c|c|}
\hline
\textbf{Clock Freq. of} & \textbf{Arrival of READ clock} & \textbf{Clock-to-Q} & \textbf{Clock-to-Q} & \textbf{Baseline} &\textbf{Proposed}  \\
\textbf{READ (GHz)} & {rel. to data in last D gate (ps)} & {baseline (ps)} & { proposed (ps)} & {works?} & {works?}\\
\hline
 30 & 2 & 8.1 & 8.1 & Yes & Yes \\
\cline{2-6}
\textbf{} & 1 & 10.6 & 8.1 & Yes & Yes\\
\cline{2-6}
\textbf{} & 0.5 & 15.2 & 8.0 & No & Yes\\
\cline{2-6}
\textbf{} & 0.1 & - & 8.1 & Yes & Yes\\
\hline
50 & 2 & 8.0 & 8.1 & Yes & Yes \\
\cline{2-6}
\textbf{} & 1 & 12.9 & 8.1 & No & Yes\\
\cline{2-6}
\textbf{} & 0.5 & - & 8.2 & Yes & Yes\\
\cline{2-6}
\textbf{} & 0.1 & - & 8.2 & Yes & Yes\\
\hline
\end{tabular}
}
\label{tab1}
\end{table}
\begin{table}[!h]
\vspace{-0.5cm}
\caption{Comparison of JJ-area of the two FIFOs}
\vspace{-0.1cm}
\centering
\tabcolsep=0.11cm
\small
\begin{tabular}{|c|c|c|}
\hline
\textbf{Number of} & \textbf{JJ-area of } & \textbf{JJ-area of}  \\
\textbf{Stages} & \textbf{baseline design ($\mu m^{2}$)} &\textbf{proposed design ($\mu m^{2}$)} \\
\hline
2 & 37.4 & 56.52 \\
\hline
5 & 118.04 & 137.16\\
\hline
10 & 255.04 & 274.16\\
\hline
20 & 523.84 & 542.96 \\
\hline
\end{tabular}
\label{tab2}
\end{table}

\section{Conclusions}
\label{sec:conc}

Challenges in low-skew global clock distribution in SFQ motivates the need for efficient CDC circuits. CDC synchronizers proposed in literature fail in high-speed test mode due to setup time violations in the capturing flip-flop. 
In this paper, we have demonstrated through simulations that multi-DRO synchronizers effectively address this problem and enable high-speed synchronization. We have also described guidelines to choose the minimum number of stages of our proposed FIFO synchronizer depending on throughput and burstiness constraints. We propose to extend this work, by thoroughly analyzing metastability in a SFQ D-element which will help better quantify its setup time. We also plan to extend the design to multi-bit synchronizers, with the use of splitter trees.

\bibliographystyle{IEEEtran}
\bibliography{IEEEabrv,bibliography}

\end{document}